\documentstyle[aps,twocolumn,epsfig,eqsecnum,rotating]{revtex}
\begin{document}

\twocolumn[\hsize\textwidth\columnwidth\hsize\csname    
@twocolumnfalse\endcsname                               

\begin{title}
{Comment on `Ising Pyrochlore Magnets: Low Temperature Properties,
``Ice Rules,'' and Beyond' by R. Siddharthan {\it et al.}	}
\end{title}

\author{
B. C. den Hertog$^1$,
M. J. P. Gingras$^{1,2}$,
Steven T. Bramwell$^{3}$, and
Mark J. Harris$^{4}$
}
\address{$^1$ Department of Physics, University of Waterloo, Ontario, Canada
 N2L 3G1                \\
$^2$ Canadian Institute for Advanced Research   \\
$^3$ Department of Chemistry, University College London, 20 Gordon
Street, London, WC1H~0AJ, United Kingdom \\
$^4$ ISIS Facility, Rutherford Appleton Laboratory, Chilton, Didcot, Oxon,
OX11~0QX, United Kingdom
}

\vspace{1mm}

\date{\today} 
\maketitle 

\vspace{-1cm}

\begin{abstract}
\end{abstract}

\vskip2pc] 

\narrowtext

Siddharthan {\it et al.}~\cite{Siddharthan} discuss the competition between
dipolar coupling and superexchange in pyrochlore magnets with
$\langle 111 \rangle$ Ising-like anisotropy, such as 
Ho$_2$Ti$_2$O$_7$~\cite{Harris1} and
Dy$_2$Ti$_2$O$_7$ \cite{Ramirez}. In the simplest
approximation for Ho$_2$Ti$_2$O$_7$, with
Ising spins and nearest-neighbor ferromagnetic exchange, one obtains the
``spin ice'' model that predicts a macroscopically degenerate
ground state~\cite{Harris1}. Consistent with this, $\mu$SR and neutron scattering 
data~\cite{Harris1,Harris2} find no phase transition in
Ho$_2$Ti$_2$O$_7$ in zero field down to at least 50 mK, and at that temperature the
neutron scattering pattern is consistent with the spin ice ground state~\cite{Harris3}.
On the other hand, the dipolar coupling, $J_D$, in these materials, 
is of the same order of magnitude as superexchange, $J_S$. Siddharthan
{\it et al.} derive an estimate for $J_S=-1.92$ K for the
near neighbour exchange and $J_D=2.35$ K.
These values are consistent with our own
estimates $J_D=2.35$K, and $J_S = -1.4\pm 0.5$ K 
from single crystal
measurements where demagnetizing effects were explicitly accounted for, and
which 
correspond to the  Curie-Weiss temperature of +1.9 K, quoted in Ref.~\cite{Harris1}.
Using their 
estimated value of $J_S$, Siddharthan {\it et al.} simulated a model
in which the dipolar summation is truncated at five nearest neighbours, and find
a  transition to a  partially ordered phase 
marked by a sharp peak in the heat capacity, $C(T)$, at $T\sim 0.8$ K~\cite{Siddharthan}.
They further argue that such a picture applies to 
Ho$_2$Ti$_2$O$_7$ by commenting that 
the  experimental heat capacity shows, as their simulation does,
a steep increase at about 1 K and that the experimental data for
Ho$_2$Ti$_2$O$_7$ suggest a vanishing entropy for the ground state
(on integrating $C(T)/T$). 
Their conclusion of a phase transition in  Ho$_2$Ti$_2$O$_7$ 
is 
inconsistent with the experimental $\mu$SR and neutron data of Refs~\cite{Harris1,Harris2,Harris3}.

In view of the very large hyperfine coupling in Ho, as observed in many other
Ho-containing compounds such as HoF$_3$~\cite{Bleaney}, it is necessary to
include a component from the nuclear spin, $I$. Bl\"{o}te {\it et al.}~\cite{Bloete}
find a heat capacity in the pyrochlore Ho$_2$GaSbO$_7$
very similar to the one of Ho$_2$Ti$_2$O$_7$, which they model below 2 K
as a simple Schottky anomaly with the theoretical maximum
for Ho ($I = 7/2$) of $0.9$  R at a temperature of about 0.3 K.

In this Comment we argue that Ho$_2$Ti$_2$O$_7$ does in fact exhibit spin ice behavior, and that the
 experimental
specific heat data
can be accounted for in terms of a {\it dipolar spin ice} model~\cite{hertog} and by including an
expected contibution from the nuclear spins~\cite{Bloete} to the appropriate long-range
treatment of the dipole-dipole interactions.

Dipole-dipole interactions are
conditionally convergent
due to their $1/r^{3}$ nature and their
lattice
summation must
be considered with care~\cite{hertog}. In order to include the 
long range nature of the dipole-dipole interaction, we have used
the standard Ewald method in our
Monte Carlo simulations with either of the above set of $\{J_S,J_D\}$ values. 
We found it sufficient to simulate 
$4\times 4\times 4$
cubic cells (1024 spins) with $\sim 10^{6}$ Monte
Carlo steps per spin. 

Figure 1 shows the electronic spin part heat capacity for $J_s=-1$ K and $J_D=2.35$ K
(open squares). The total entropy is within 2\% of
$R\{\ln(2)-(1/2)\ln(3/2)\}$ and therefore agrees with that of spin ice~\cite{Ramirez,hertog}.
We have found in our Ewald simulations 
that dipolar spin ice occurs for  all values $J_S/J_D > -0.91$, which
includes the value of Ref.~\cite{Siddharthan}, and that fully developped $Q=0$
long-range antiferromagnetic order occurs for  $J_S/J_D < -0.91$, with no
partially  frozen or ordered state as found in Ref.~\cite{Siddharthan}.
The open circles
show the sum of this magnetic heat capacity with a 
nuclear Schottky anomaly as found in Ho$_2$GaSbO$_7$~\cite{Bloete}.
We find that these results
reproduce reasonably well the
experimental data presented in Ref.~\cite{Siddharthan} (filled circles),
and is definitely
more accurate than the model proposed by Siddharthan {\it et al.}
Equivalently, subtracting the nuclear contribution from the magnetic
heat capacity  measured in Ref.~\cite{Siddharthan} leads to a
heat capacity similar to that found in the spin ice material
Dy$_2$Ti$_2$O$_7$~\cite{Ramirez}.
In conclusion, we argue that the experimental specific heat of
Ho$_2$Ti$_2$O$_7$ can be accounted for by the sum of a {\it dipolar spin ice}
magnetic contribution and a nuclear hyperfine Shottky anomaly with no
need to invoke a transition to a partially ordered state.
\begin{figure}[h]
\begin{center}
  {
  \begin{turn}{90}%
    {\epsfig{file=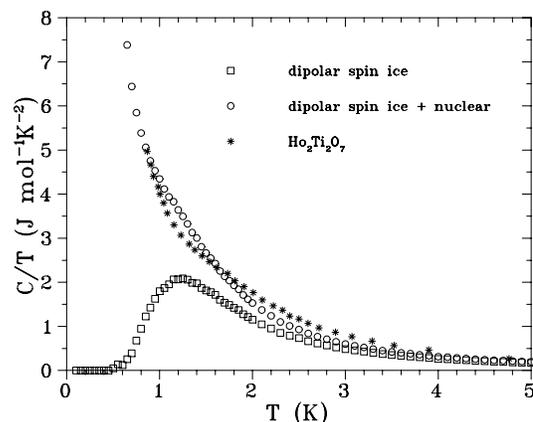,width=5.5cm} }
   \end{turn}
   }
\caption{Open squares: heat capacity of dipolar spin ice with
$J_s=-1$ K and $J_D=2.35$ K.
Open circles: same data with the additional nuclear contribution
expected for Ho [6,7].
Stars: experimental data extracted from Ref. [1].}
\end{center}
\end{figure}

\end{document}